\documentstyle[psfig,11pt]{article}

\begin{document}

\centerline{\large\bf A DIFFERENTIAL METHOD OF SEARCH FOR}
\vskip 2mm
\centerline{\large\bf THE CMBR SPECTRAL-SPATIAL FLUCTUATIONS}

\bigskip

\centerline{\bf Victor K. Dubrovich$^1$, Anisa T. Bajkova$^2$}

\bigskip

\centerline{$^1$\it Special Astrophysical Observatory RAS, e-mail: dubr@MD1381.spb.edu}

\medskip

\centerline{$^2$\it Institute of Applied Astronomy RAS, e-mail: bajkova@quasar.ipa.nw.ru}

\bigskip
\begin{abstract}
The CMBR spectral-spatial fluctuations (SSF) formed in early Universe
during the Dark Ages are considered. Main attention is focused on the
narrow-band spectral properties of the SSF. Based on these properties we
propose to use a differential method in order to search for these fluctuations.
Description of the method is given.
\end{abstract}

\section{Introduction:  Structure Formation and Big Bang Cosmology}

        The Big Bang theory for the expanding Universe is now well
established. One of the main observation effect caused by this expansion
is the redshift of photons, expressed as $\lambda_{obs}=(z +1)\cdot
\lambda_{em}$,
where $\lambda_{em}$ is the emitted or rest wavelength of the photon, and $z$
is the redshift.

   One early moment of particularly great
importance occurs at a redshift $z\approx 1100$, corresponding to a time
$3\cdot 10^5$ yr.
after the Big Bang. This is the moment observed when we study the
distribution of the cosmic microwave background radiation (CMBR), heat left
over from the Big Bang. The importance of this surface is explored in reviews
such as Silk, 1994. In effect, the distribution of intensity of CMBR
on the sky provides information about the large-scale distribution
of matter at a very early time in the history of the Universe. For this
reason, both NASA and the European Space Agency have attached high priority
to the mapping of this surface using the CMBR. We know from earlier work
that it is largely featureless, that is that the distribution of matter was
quite homogeneous early in the history of the Universe. However, the COBE
satellite (Bennett et al., 1993), and recently BOOMERANG (Lange et al.,
2001) and MAXIMA (Hanany et al., 2000) revealed
small fluctuations in the CMBR temperature, corresponding to perturbations
in the density of matter of order $10^{-3}$. These are the seeds of all the
structure we see in the Universe today. To summarize, the Big Bang itself
was quite homogeneous, but very low amplitude fluctuations in density were
present at an epoch of $3\cdot 10^5$ yr.

On the other hand, the present Universe is exceedingly lumpy~ ---~ there
are large-scale structures ranging in size from galaxies to clusters of
galaxies and even larger systems. Recent optical and radio studies, for
instance by the Hubble Space Telescope, have established that structure is
present in the Universe back to redshifts at least $z=5$.

It follows from these two sets of observations that much of the
structure currently seen in the Universe first formed in the redshift
interval 6--1000. It is a deeply interesting question to know when and how
the large-scale systems we see in the Universe formed. Unfortunately, there
are essentially no observations available in that redshift interval,
corresponding to the time period $3\cdot 10^5$ to $10^9$ years after the Big
Bang (although there are some hints from Lanzetta et al., 1999 of the existence
of galaxies at $z>6$). Perhaps for this reason, this phase in the history
of the Universe is known as the "Dark Ages" of the Universe.

\section{The Dark Ages}

        While we have no direct observations of objects or proto-objects in
the Dark Ages, we may make some inferences. First, sometime in that
interval, the small amplitude, linear perturbations in the density of
matter must have become nonlinear, since the densities within galaxies
observed even at $z=5$ are much higher than the background density. The
physics of nonlinear gravitational collapse are much more complicated and
rich than the simpler physics of linear gravitational contraction. In other
words, a good deal of interesting astrophysics certainly occurred during
the Dark Ages. Also, some very approximate constraints can be placed on
events during the Dark Ages by looking at integrated background light. For
instance, there are upper limits on the energy release at that range of
redshifts established by observations by the COBE satellite (see Hauser et
al., 1998; Dwek et al., 1998 and Haarsma and Partridge, 1998).

        Given the importance of questions about the formation of structure
in our Universe, it is obviously important to explore any means to increase
our knowledge about the Dark Ages of the Universe. The interaction of the
CMBR with molecular species created in the early Universe offers such a
possibility. In particular, the interaction of the CMBR with lumpy matter
containing simple molecules will introduce wavelength dependent
anisotropies in the CMBR:  these are the spectral-spatial fluctuations
(SSF) described in detail by Dubrovich (Dubrovich, 1977, 1994, 1997). The
amplitude of the SSF signals is not
large, but current and planned observation programs offer the possibility
that these clues to the formation of structure early in the Universe can be
detected.

\section{Theory of Spectral-Spatial Fluctuations}

        During the Dark Ages, there is a period when the material contents
of the Universe are present in the form of atoms and molecules, with low
amplitude density fluctuations and a low number of free electrons. The main
"objects" at this epoch are diffuse and low-contrast protogalaxies. The
temperatures of matter and radiation (the CMBR) are practically equal until
the redshift drops to 150.

It is possible to generate energy within these proto-objects as a result of
heating due to their collapse and consequent radiative cooling. But the low
matter density and low ionization lead to very low luminosity in free-free
emission or atomic transitions. Another mechanism which can produce
observation effects is the Doppler effect by peculiar motion of
a protogalaxy with a velocity $V_p$ (Dubrovich, 1977, 1997; Maoli et
al., 1996).
Note that in this case there is no need for an
internal energy source in the proto-object. The value of the distortion
produced in the CMBR is:
\begin{equation}
\delta T/T = (V_p/c) \tau,
\end{equation}
where $c$ is the speed of light and $\tau$ is the
optical depth of the proto-object ($\tau<1$). In the case of resonant transitions (line
emission) in molecules, the optical depth $\tau$ can be strongly frequency
dependent, having a high value in a narrow bandwidth around each resonant
frequency.
But in any cases it is possible to make one general affirmation:
due to this mechanism one can obtain only the value
$\delta T/T<V_p/c<10^{-3}$.

        Several authors have investigated the most probable primordial
molecules and the effects they create. The expected abundance of these
simple molecules in the early Universe was discussed by Puy et al., 1993;
Lepp and Shull, 1984; Palla et al., 1995; Maoli et al., 1996 and Stancil et
al., 1996.

        We now show that these primordial molecules, which may produce
observable effects in the CMBR, do trace the initial large-scale
distribution of matter (1). The rate of molecule formation depends on many
physical parameters such as the electron density, $n_e$. Free electrons
catalyze the formation of neutral molecules, for instance $H_2$; in turn, the
abundance of other molecules are strongly correlated with $H_2$ concentration.
It follows that fluctuations in the matter density and $n_e$ lead to nonlinear
enhancement of the optical depth in molecular lines.
   The kinetic temperature of matter also is involved after $z\approx 150$. It is
higher in the denser regions, leading to a shift of equilibrium in the ratio
of $H^{+}/H_2$, $HeH^{+}/H_2$ and $LiH/H_2$
towards higher values (2). However, the peculiar velocity continues to be
the dominant effect, and this results from hydrodynamic motions of matter
strongly correlated with density fluctuations (e.g., Vishniac, 1987).
Later, nonlinear stages of the evolution of proto-objects are characterized
by the high temperature of matter and high ionization due to compression
and shocks. This lies in the interval $50>z>5$. Ionizing photon flux may
be sufficient to ionize hydrogen, but not strong enough to render the
protogalaxies visible in the optical or the radio range. The Doppler
mechanism still dominates. The $He H^{+}$ molecule will play a particularly
important role. The rest wavelength of its first rotational transition is
0.149 mm, with other lines at 0.075 mm and 0.05 mm. At $z=5$, these become
0.9 mm, 0.45 mm and 0.3 mm, respectively.

        We note that $He H^{+}$ will be present wherever high temperature,
compressed matter is present. In addition, the fact that the effect on the
CMBR is a resonant one with $t$ depending strongly on frequency means that
observations at any fixed wavelength are strongly dependent on the
redshift. Thus we can obtain a complete 3-dimensional picture of structural
formation including the radial dimension, practically impossible to obtain
in any other way.

        The final stage of the evolution of matter is the formation of the
first stars. The starburst phenomena will lead to the synthesis of heavy
elements, especially $C, N$ and $O$. This allows for the formation of molecules
such as $OH^{+}, OH, CH$ and $CH^{+}$, etc. At the same time, some matter will
be accelerated to high speeds due to the expulsion of shells in the starburst
phenomena. We will investigate whether these molecules also produce
observation SSF effects.

\section{Spectral properties of SSF}

Except of pure scattering the luminescent transformation of some kind of
superequilibrium photons is possible as well as the combination of both
mechanisms (Dubrovich, 1977, 1994, 1997, 1997a, 1999, Dubrovich et al., 1995).
In all these cases the essential peculiarity of the fluctuations, namely
strong frequency dependence of the effect, becomes apparent.
So, if an object contains gas of any molecules and has a redshift
$z$, it is seen only at frequencies $\nu_i$:
\begin{equation}
\nu_i = \nu_{oi}/(1+z),
\end{equation}
where $\nu_{oi}$ is a discrete set of the molecule's rest transition frequencies.
It means that an object can be observed at different but discrete
frequencies. And vice versa, if some object is seen at the given
frequency $\nu$, it means that this object can have any of the redshifts $z_i$:
\begin{equation}
z_i = \nu_{oi}/\nu - 1.
\end{equation}

This uncertainty of $z_i$ can be cleared up, for instance, by such a way as for
distant quasars: it is necessary to obtain spectrum of interested object
wide enough. If we suppose that this line caused by any rotational transition
in simple two-atomic molecules it is necessary to get the spectrum interval
from $\nu_1$ to $\nu_2=2\cdot \nu_1$. In this case we can guarantee, that all
possible
variants of rotational number $J=0,1,2,...$ corresponded to this transition could be
known.
Indeed, the frequency of rotational transition (from level with rotational
number $J$ to level $J+1$, for scattering) $\nu_J$ is proportional to
$J+1$. So the
frequencies ratio of two nearest observed lines will be $(J+2)/(J+1)\le 2$. The
maximum of this ratio corresponds to the lines from $J=0$ to $J'=1$ and
from $J'=1$
to $J''=2$. If we see the ratio less then 2 we can calculate $J$ which
is the number of the low rotational level. For example, if this ratio is
1.5 it means that we see transitions from $J=1$ to $J'=2$ and from
$J'=2$ to $J''=3$.

One of the consequences of (2), (3) is that the point object with redshift
$z_1$ visible at frequency $\nu_1$ is not visible at another frequency
$\nu_2$, if the following relation is realized:
\begin{equation}
\nu_1<\nu_2;~~\nu_1 =\nu_{oi}/(1+z_1);~~\nu_2<\nu_{o(i+1)}/(1+z_1).
\end{equation}

Here we suppose that the size $L$ of the object is proper small.
More exact condition for $\Delta\nu$, taking into account the finite
size of an object, will be done below.
At the same time at the frequency $\nu_2$ we will see the objects with
redshift $z_2$ in accordance with (3), which may be different from the
seen at the frequency $\nu_1$.

The angular size $\theta$ and the frequency bandwidth $\Delta\nu$, in which
the given object can be observed, depends on its linear size $L$.
For small values of $z$ the angular size of an object (when the linear
size is constant and equal to $L$) is decreasing with moving away,
but for $z$ larger than some value, it is increasing.
The modern observation data select the model of plane expanding Universe
with cosmological constant $\Lambda$.
Exact relations between $\theta$ and $z$ in this model contain the
parameters $\Omega_m$ and $\Omega_{\Lambda}$, which are the ratios of the
matter $n_m$ and "vacuum" densities to the critical density $n_c$ respectively
($\Omega_m+\Omega_{\Lambda}=1$) and look like (Sahni et al., 1999):
$$
\theta=(H_0L/c)(1+z)/\phi(z),
$$
$$
\phi(z)=\int_0^z[\Omega_m(1+z)^3+\Omega_{\Lambda}]^{-1/2}dz.
$$

For large $z$ and $H_0=60$km/s$\cdot$Œps$\cdot$$h_{60}$,
$\Omega_m=0.3$ and $L$ in Mps we obtain
$$
\theta \approx 15^{\prime\prime}h_{60}L(1+z).
$$

On the other hand, front and back edges of molecular cloud are at
different distances from us, i.e. have different $z$.
The value of this difference of $z$ is connected with $L$
in the following way (Sahni et al., 1999):
$$
\Delta z/(1+z) = (H_0 L/c) [\Omega_m(1+z)^3+\Omega_{\Lambda}]^{1/2}.
$$

The fact that the cloud occupies the interval of redshifts $z$
means, that if it radiates or reflects the radiation locally in sufficiently
narrow lines, then all the radiation occupies the frequency interval
$\Delta\nu$, moreover
$$
\Delta\nu/\nu=\Delta z/(1+z).
$$
Consequently,
$$
(\Delta\nu/\nu)\approx \theta\phi(z) \cdot [\Omega_m(1+z)^3+\Omega_{\Lambda}]^{1/2}(1+z)^{-1}.
$$
Numerically for large $z$ and $\Omega_m=0.3$ we obtain
\begin{equation}
(\Delta\nu/\nu)\approx 1.310^{-3}(\theta/1^{\prime}) \cdot [\phi(z)/\phi(20)](1+z)^{1/2}.
\end{equation}

If we simultaneously observe with two receivers with frequency
difference $\Delta\nu$, the fluctuations with the size larger than $\theta$
are seen at the both. If the size of the fluctuations is smaller, they
are seen only with one of the receivers, another receiver will detect
only a background noise (Dubrovich, 1982).
Thus the correlation function of these two observations will be
equal to within a noise to zero at small scales and different of zero
at scales greater than certain.
The presence of clear break is possible only if all the objects are
of equal size $L$.
Practically there is some spectrum of scales.
The effect considered shows that the initial spectrum can be
significantly distorted.
More full investigation of this problem stands out of this paper.
But from the said already it is seen that the analysis of SSF will
give additional information on the matter parameters and redshifts $z$
of basic proto-objects.

Scattering by molecules in several lines leads to superposition of
images of different objects of different $z$.
As shown in Dubrovich (1977), the cosmological molecules $LiH$ and $HD^{+}$
can exist at $50<z<200$.
The interval of wavelengths for $LiH$ spectral lines in this condition
lies between 13 cm for $z=200$ and $i=1$ and $0.5$ cm for $z=50$ and $i=6$.
For $HD^{+}$ it spreads from 4.4 cm to 0.2 cm.
This leads to increasing of fluctuations number at large frequencies.
The power spectra obtained in cm and mm ranges must differ by the increase
of the part of small scale fluctuations in the second case.

The estimations and equations mentioned above are correct for objects
freely expanding over the Universe in accordance with the Hubble law.
But really at the sufficiently late stages the deceleration of their
expansion takes place and contraction begins due to self-gravitation.
In this case the line's width can be considerably smaller, and the
amplitude larger (Zel'dovich, 1987).

\section{The differential method for detecting the SSF}

The differential method can be considered as an alternative method to the
correlation function analysis indicated above.
In this case the subject of interest is actually the first derivative
of spatial distortions with respect to the frequency.

The method used based on the analysis of a difference of two CMBR
temperature maps
observed at different frequencies and reduced to one beam shape.
Obviously, such a difference map contains information only on the
secondary fluctuations because the primary CMBR fluctuations present in
both maps will be eliminated in subtraction result due to their black-body
spectrum nature.

Let the difference of the CMBR observation frequencies be equal to
$\Delta\nu_1$.
Let the limiting angular fluctuation size $\theta_1$ correspond to this
frequency difference in accordance with equation (5).
Obviously, the fluctuations of size larger than $\theta_1$ will be seen
at both maps. After subtraction of one map from the other the
fluctuations of size larger than $\theta_1$ will be mutually compensated
and the remainder map will contain only the fluctuations smaller
than $\theta_1$.
Evidently, this fact can be seen also from angular power spectrum of
this map which is close to zero at frequencies with multipoles $l<l_1$
and different from zero at $l>l_1$, where $l_1\approx 1/\theta_1$ is the
break of the spectrum.
Now let the difference of two frequencies be $\Delta \nu_2$, where
$\Delta \nu_2 > \Delta \nu_1$.
The angular size $\theta_2$ corresponds to this frequency interval.
Obviously, $\theta_2 > \theta_1$ in accordance with equation (5).
In this case we will see at the difference map the fluctuations with
size smaller than $\theta_2$ and the corresponding angular power spectrum
will have the break at $l_2\approx 1/\theta_2 <l_1$.

In Figs.~1 and 2 the illustration of the differential method is given.
In the upper row four maps (1, 2, 3 ¨ 4) observed at four different
frequencies $\nu_1, \nu_2, \nu_3, \nu_4$ are shown.
All the maps have identical angular power spectrum depicted in Fig.~2 by
bold dashed line.
The maps shown differ from each other in the following way. Spectrum
harmonics of maps with number 1 and number $i (i=2,3,4)$ are
characterized by identical phases for $l\le l_s(i)$ (in our case
$l_s(2)=640, l_s(3)=1152, l_s(4)=1536)$ and different ones for
$l>l_s(i)$. Note that the phases are distributed randomly in interval
$[0,2\pi]$.

In the bottom row of Fig.~1 three difference maps indicated as 1-2, 1-3
and 1-4 obtained after subtraction from the map with number 1
of maps with numbers 2, 3 and 4 correspondingly.
The frequency differences satisfy the following inequality
$\nu_1-\nu_2>\nu_1-\nu_3>\nu_1-\nu_4$.
Respectively, maximal size of the CMBR fluctuations visible on the
remainder maps meets the inequality: $\theta_{1-2}>\theta_{1-3}>\theta_{1-4}$.
Qualitatively, the scale of the fluctuations can be estimated directly
from the maps shown in Fig.~1. The angular power spectra of difference maps
1-2, 1-3 and 1-4 are depicted in Fig.~2 by solid, dotted and thin dashed
lines correspondingly.
As seen the break points of the angular power spectra meet the following
relation $l_{1-2}<l_{1-3}<l_{1-4}$. Obviously,
$l_s(2)=l_{1-2}, l_s(3)=l_{1-3}, l_s(4)=l_{1-4}$.

In the example considered above it was assumed that the observed CMBR maps
consist only of the primary and secondary CMBR fluctuations.
But in practice the CMBR detected maps contain in addition a number of
foregrounds such as pixel noise, extragalactic unresolved sources,
free-free, synchrotron radiation, Sunyaev-Zel'dovich effect, Galactic
dust and so on.
The most dangerous foreground for the differential method considered is
the pixel noise. Other foregrounds characterized by considerably
low value of spectral index as compared with analyzed secondary CMBR
fluctuations and are successfully eliminated in resulted remainder maps.
Below we demonstrate the simulation results of the differential method
in presence of some additive pixel noise in the CMBR maps.

For this Gaussian zero mean white noise was added to the CMBR
maps 1,2,3 and 4. Signal-to-noise ratio for each map is $\approx 1.44$.
Noisy maps 1,2,3 and 4 are shown in the upper row of Fig.~3.
The difference maps 1-2, 1-3, 1-4, obtained after lower-frequency spatial
filtering with bandpass equal to the CMBR spectrum width, are shown in
the lower row of Fig.~3. As seen from the pictures the presence of the
noise in low-frequency region of the CMBR power spectrum led to
appreciable mess of fine structure of the remainder fluctuations.
Therefore in order to obtain the necessary both qualitative and
quantitative characteristics the analysis of the angular power spectra
of the difference CMBR maps is required.
The angular power spectra of difference maps
1-2, 1-3 and 1-4 are depicted in Fig.~4 by solid, dotted and thin dashed
lines correspondingly.
As seen from the pictures, the angular power spectrum breaks corresponding
to boundaries between the noise spectrum and spectrum of CMBR+noise map
are quite apparent even without preliminary processing of the detected
CMBR maps. 
Obviously, in case of higher level of noise a more complicated preliminary
signal processing directed to decreasing the noise in the CMBR spectrum
region would be required (Bajkova et al., 2002).

Availability of the sufficiently accurate measured value of the CMBR angular
power spectrum break $l_c$ allows us to estimate $\theta_c$, what in
turn allows to determine the retardation parameter $q_0$, redshift $z$,
and type of molecules of cosmological origin in accordance with the
method explicitly described in (Dubrovich, 1982).

\section{Conclusion}

Now we formulate the basic results. Presence of molecular clouds in
the early stage of the Universe evolution $(z>100)$ must lead to
forming spectral-spatial fluctuations of the CMBR.
The amplitude of these fluctuations can achieve $310^{-3}K$ in the case
of simple scattering and exceed this value in case of sufficiently
powerful early extraction of energy what allows to state a problem
about their explicit investigation.
As shown in this paper, the use of the correlation analysis (Dubrovich,
1982) or
considered here more reliable differential method principally allows us
both to detect the difference of the primary molecules and determine the
type of each of them. Moreover the determination of the interval of $z$,
in which these molecules exist as well as parameter $\Omega_m$ is possible.

\begin{figure}

\centerline{
\psfig{figure=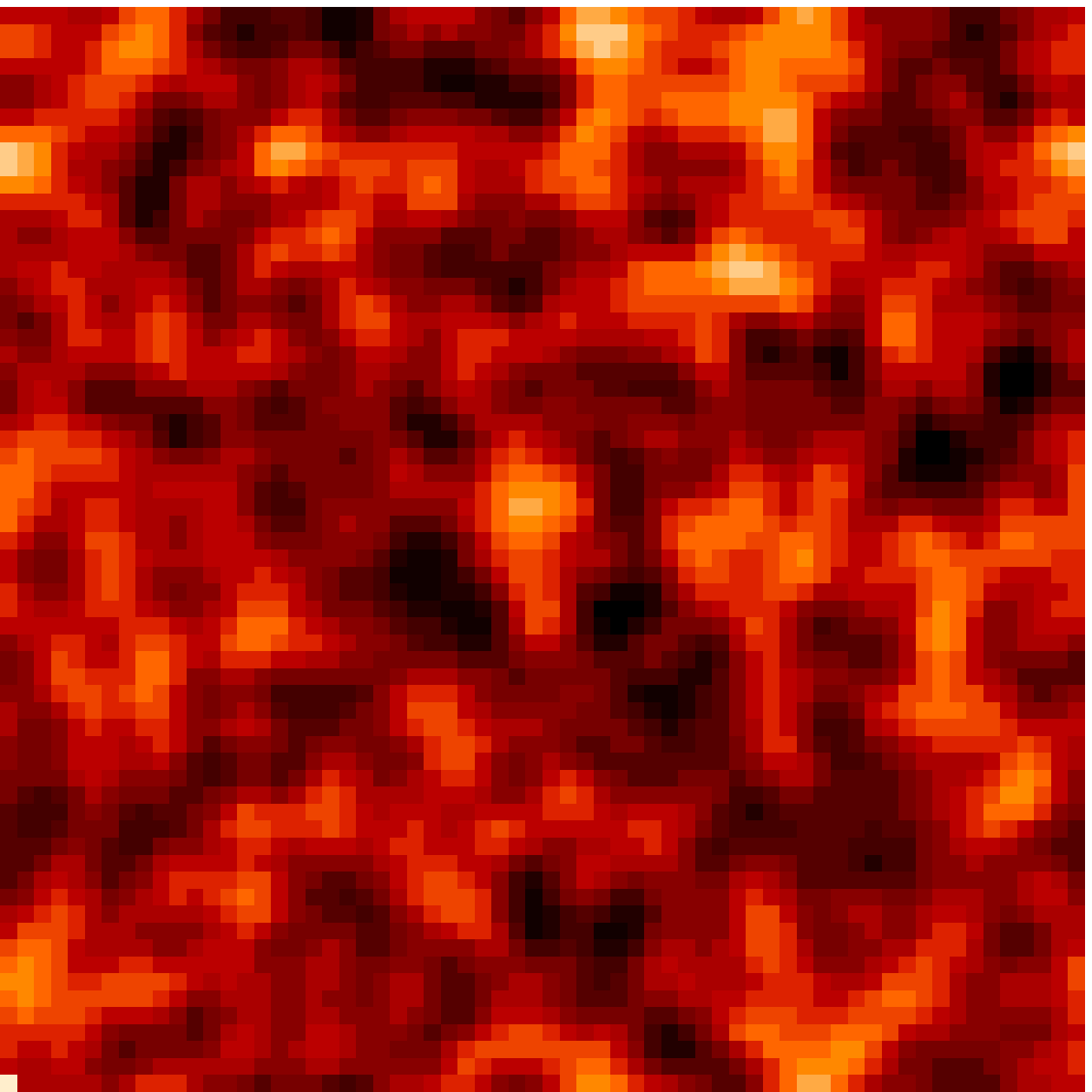,width=50mm}
\hspace{-10mm}
\psfig{figure=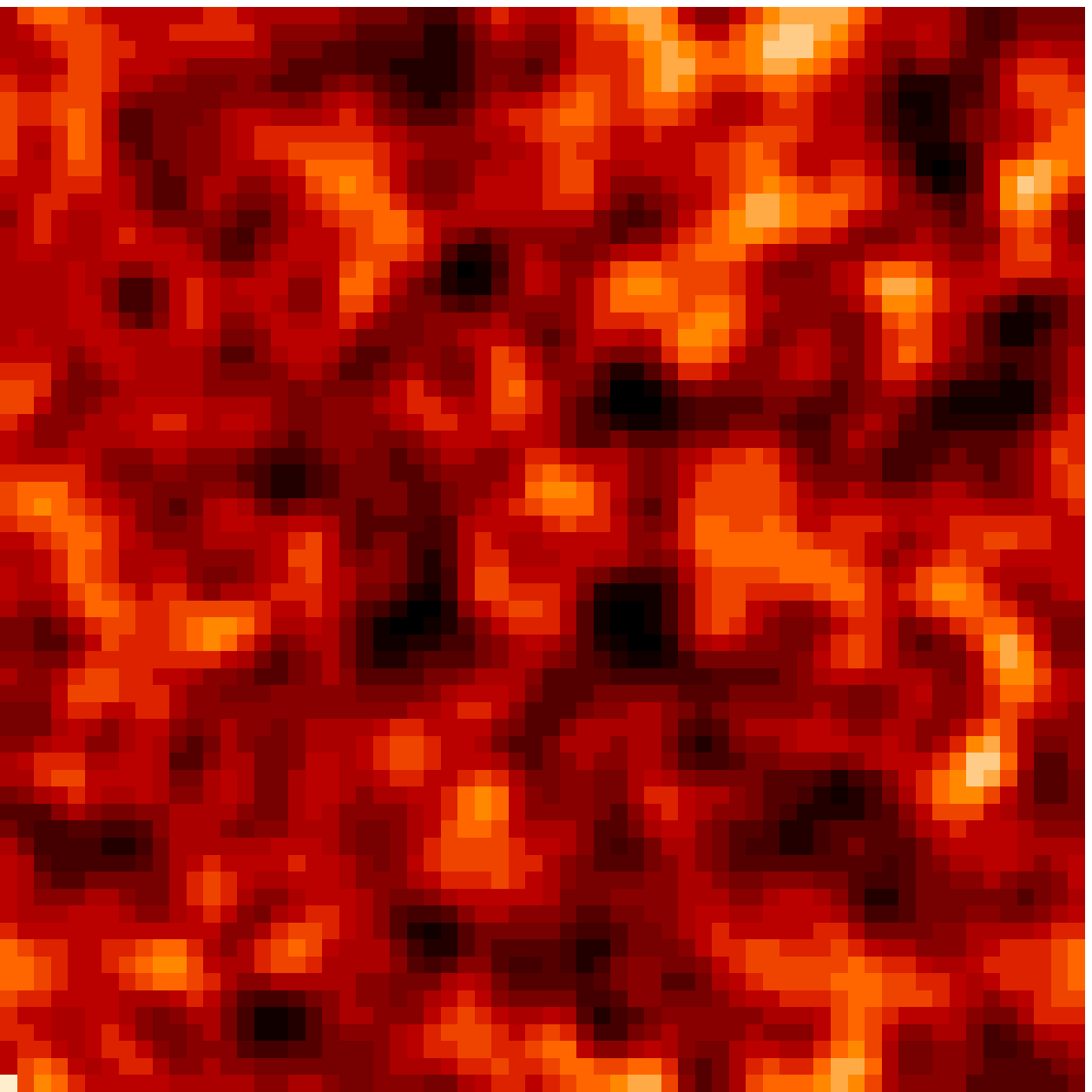,width=50mm}
\hspace{-10mm}
\psfig{figure=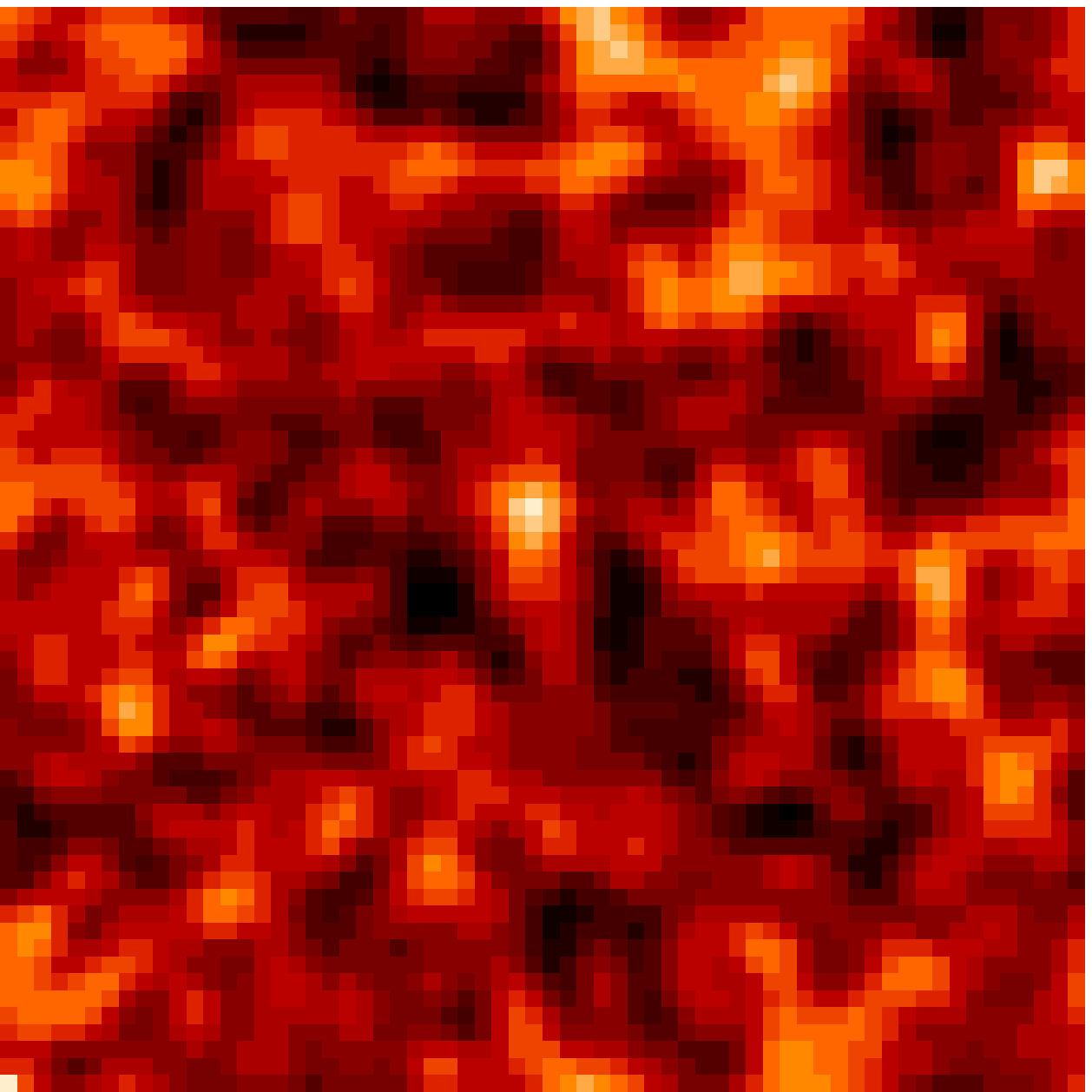,width=50mm}
\hspace{-10mm}
\psfig{figure=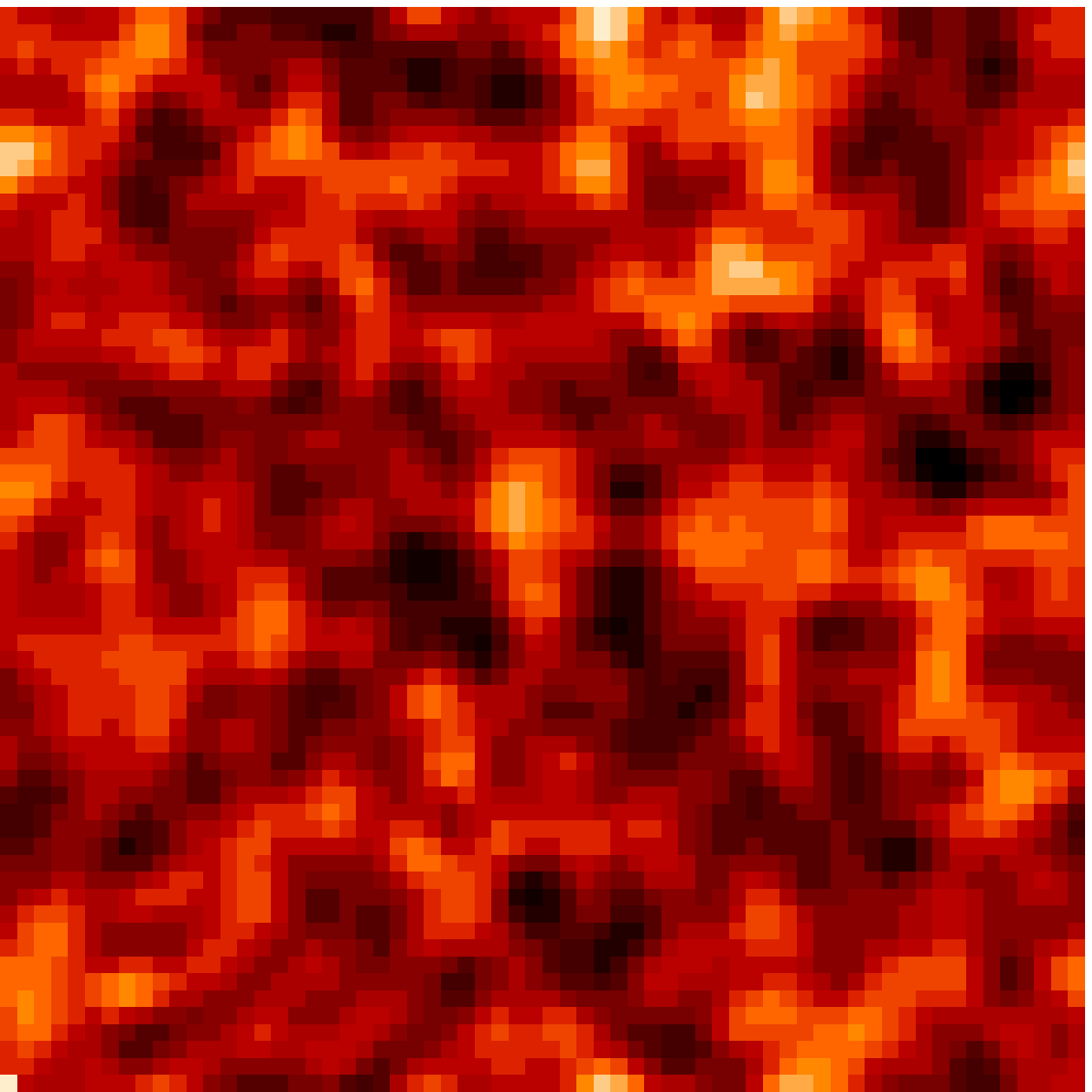,width=50mm}
}
{\hskip 11mm 1 \hskip 39mm 2 \hskip 39mm 3 \hskip 39mm 4}

\vskip 5mm

\centerline{
\psfig{figure=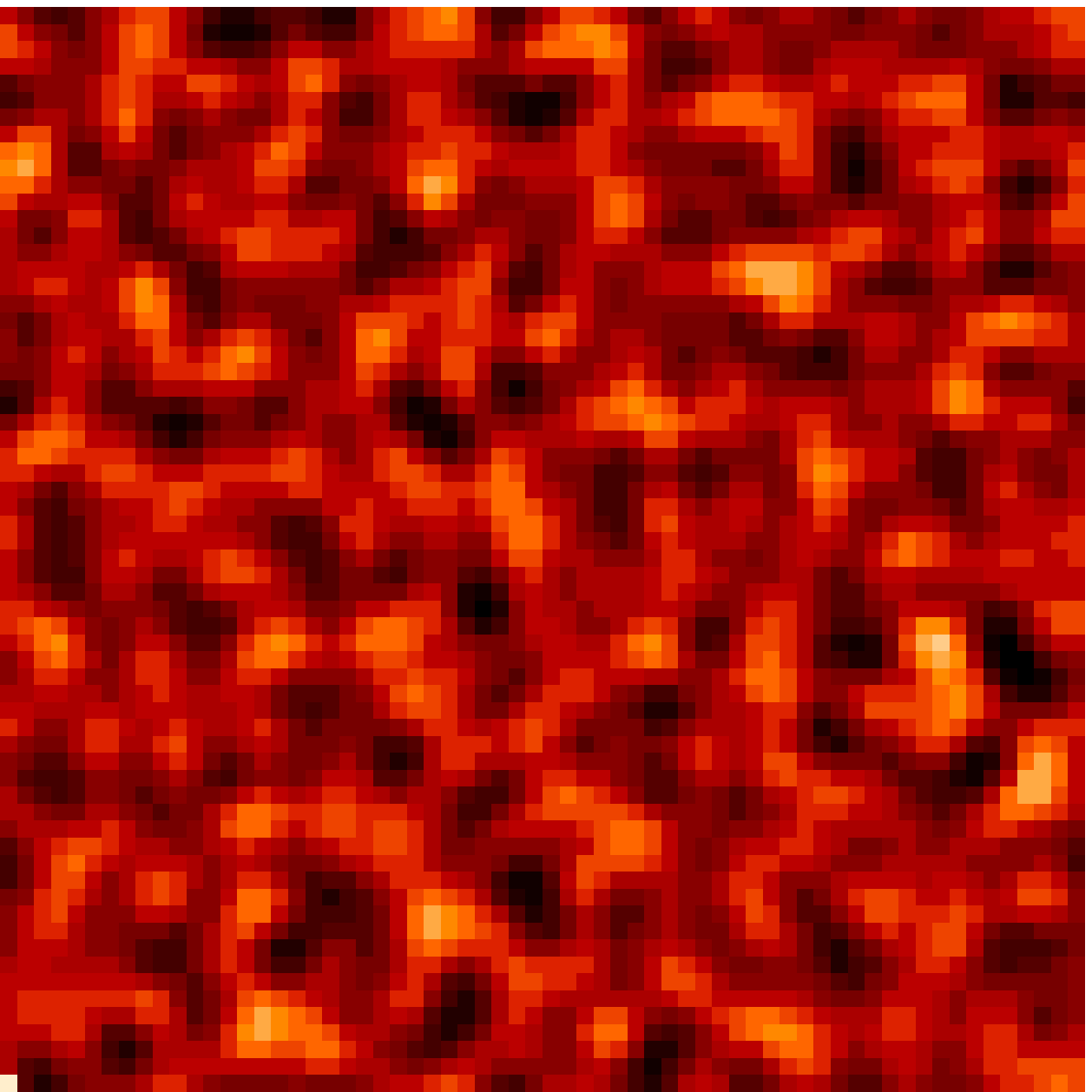,width=50mm}
\hspace{11.5mm}
\psfig{figure=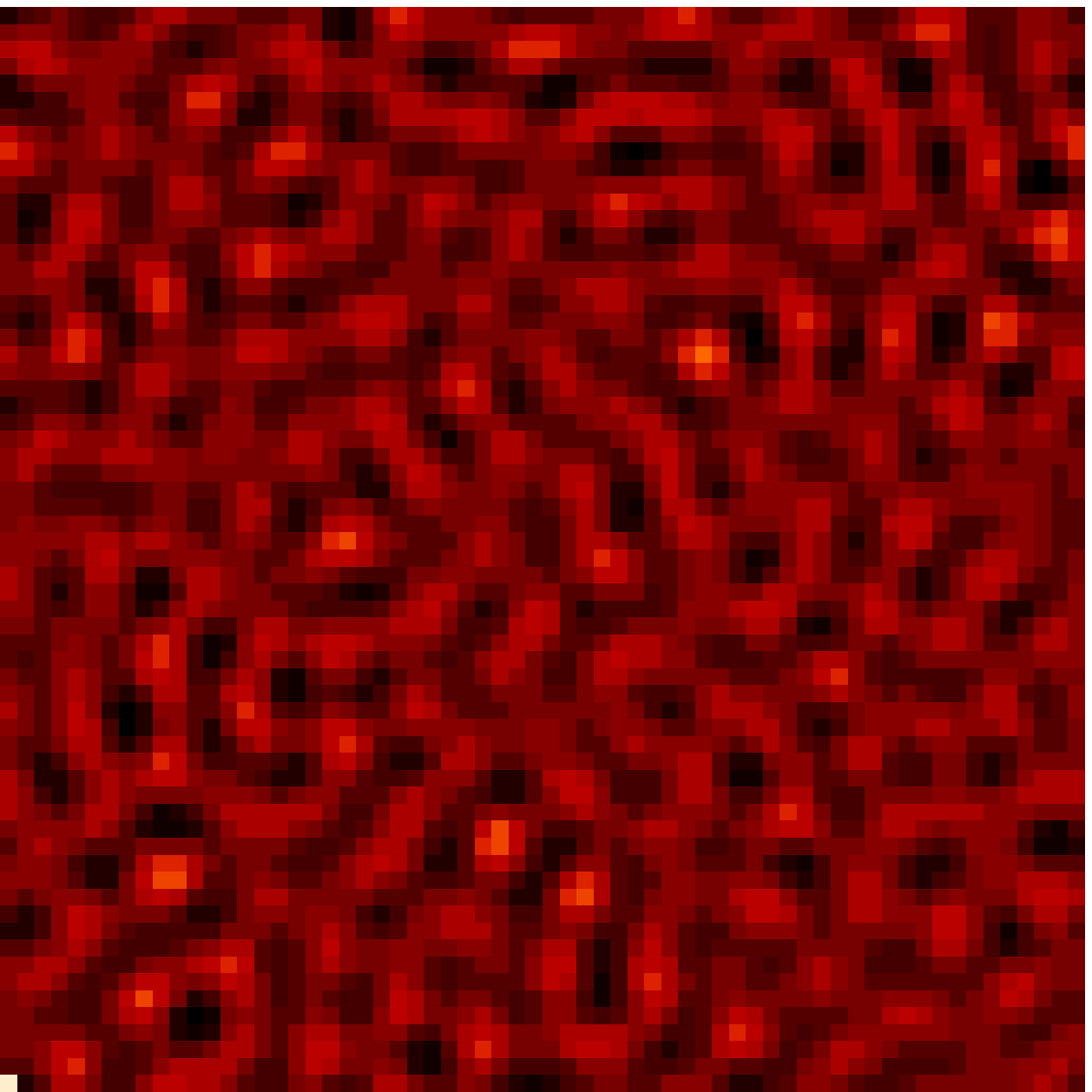,width=50mm}
\hspace{11.5mm}
\psfig{figure=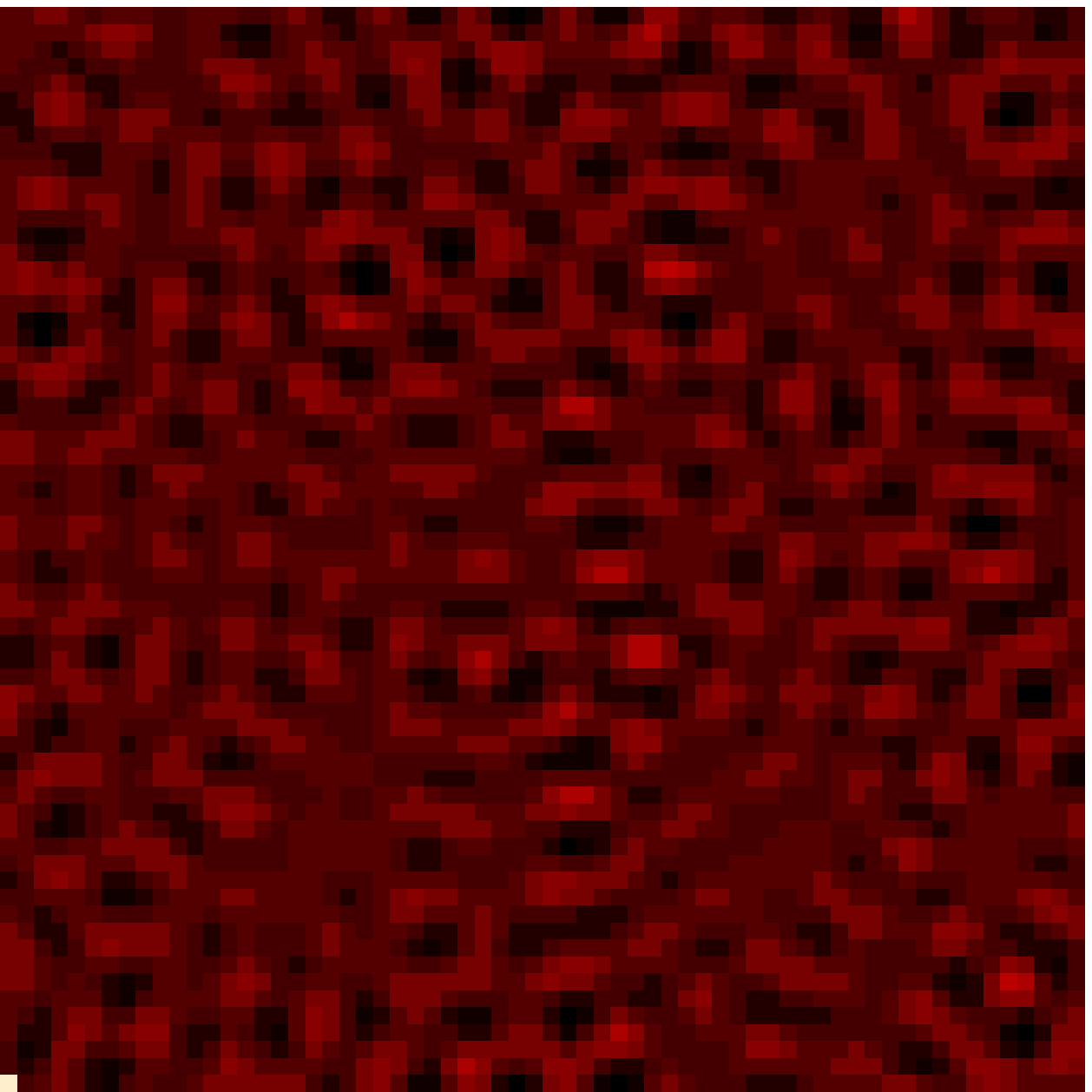,width=50mm}
}
{\hskip 9mm 1-2 \hskip 57mm 1-3 \hskip 58mm 1-4}
%\caption{}
\vskip 5mm
\centerline{\bf Fig.~1}
\end{figure}

\begin{figure}

\centerline{
\psfig{figure=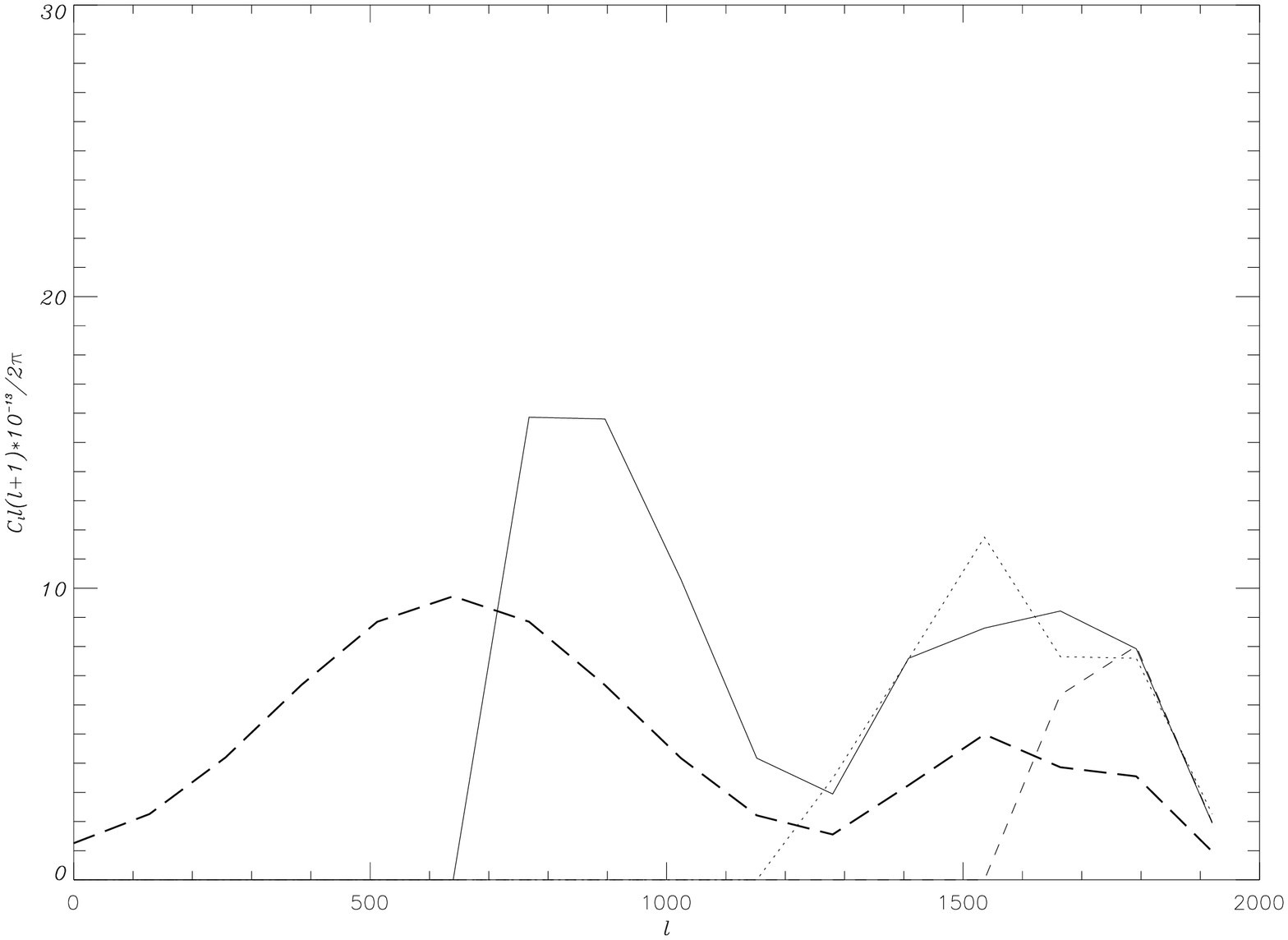,angle=90,width=150mm}
}
%\caption{}
\medskip
\centerline{\bf Fig.~2}
\end{figure}

\begin{figure}

\centerline{
\psfig{figure=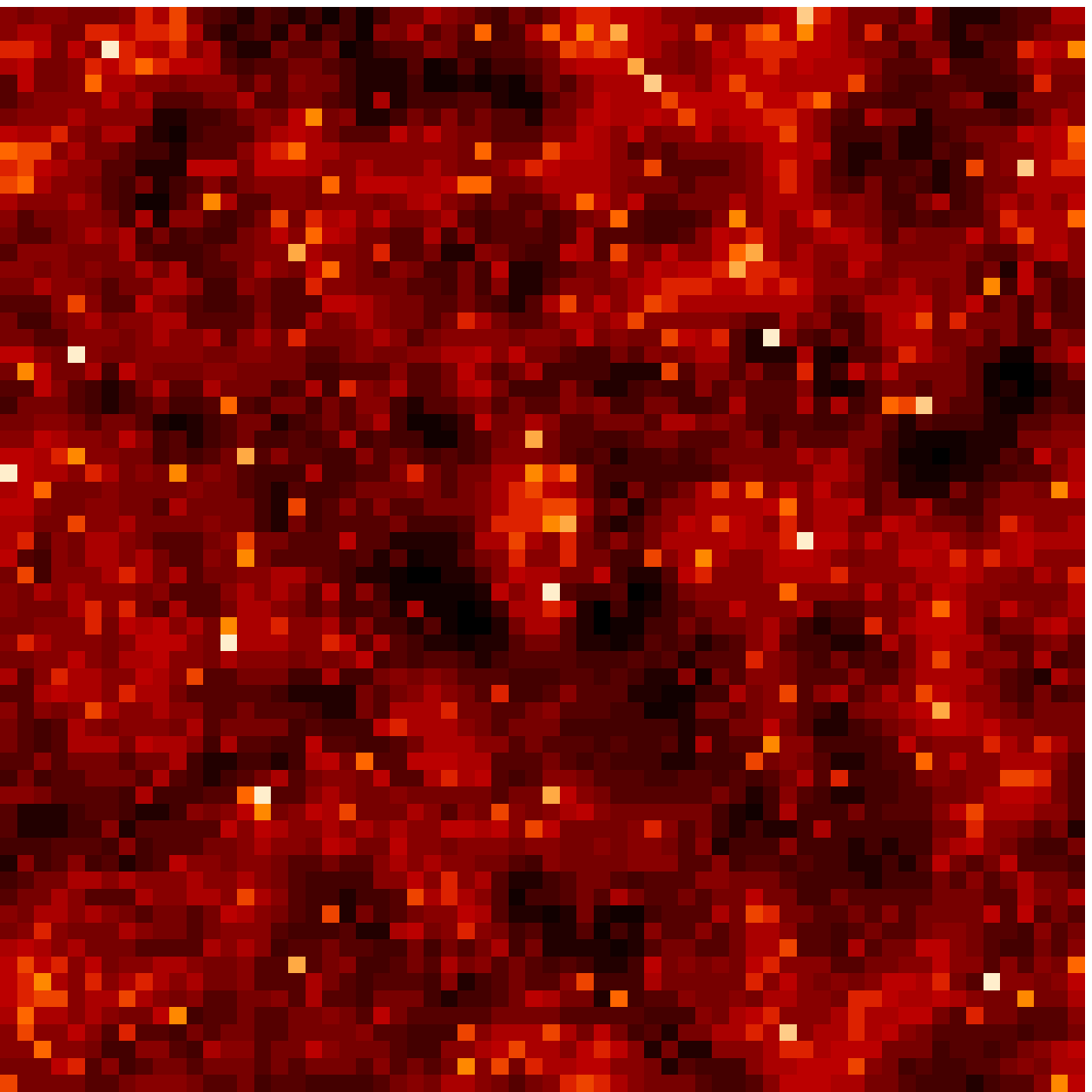,width=50mm}
\hspace{-10mm}
\psfig{figure=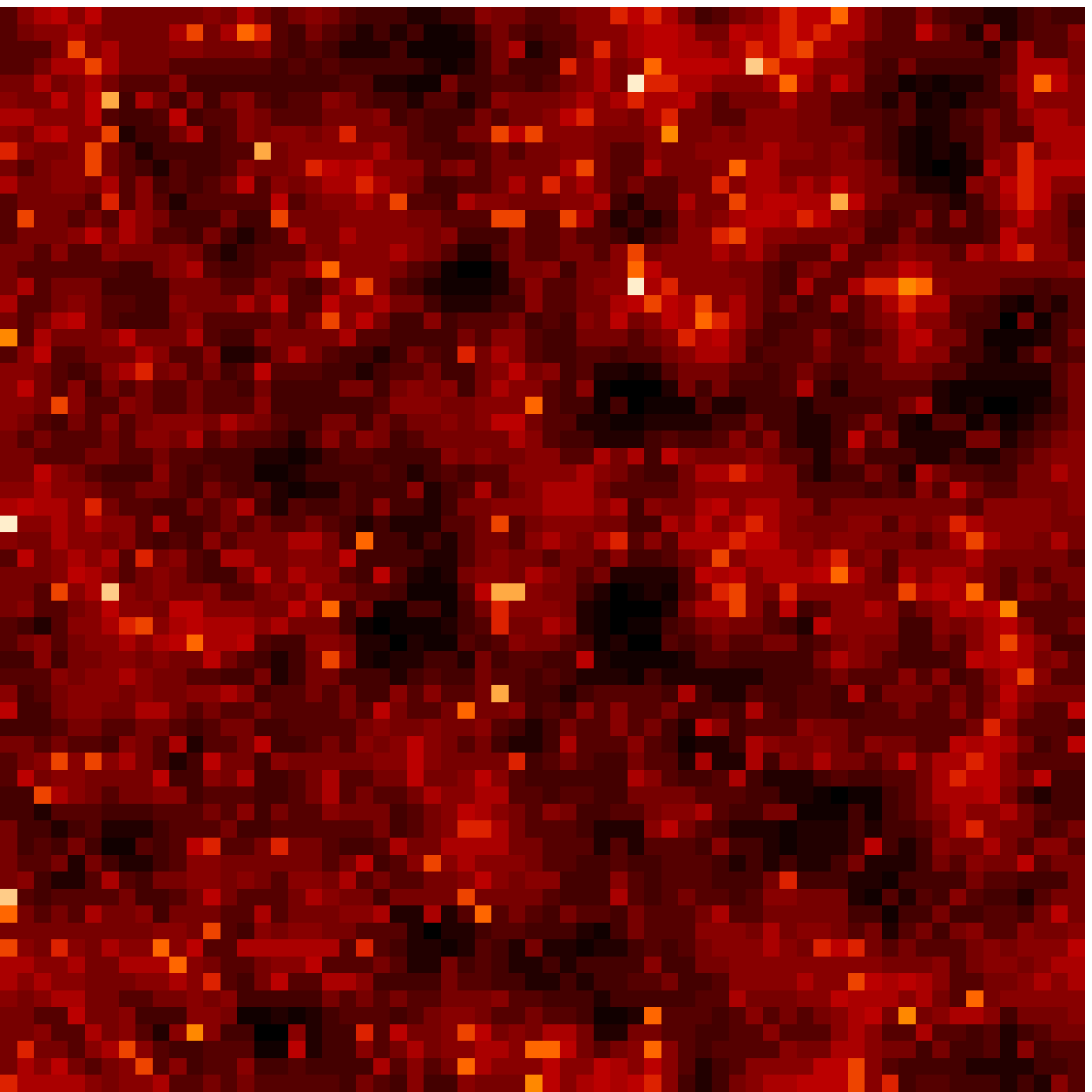,width=50mm}
\hspace{-10mm}
\psfig{figure=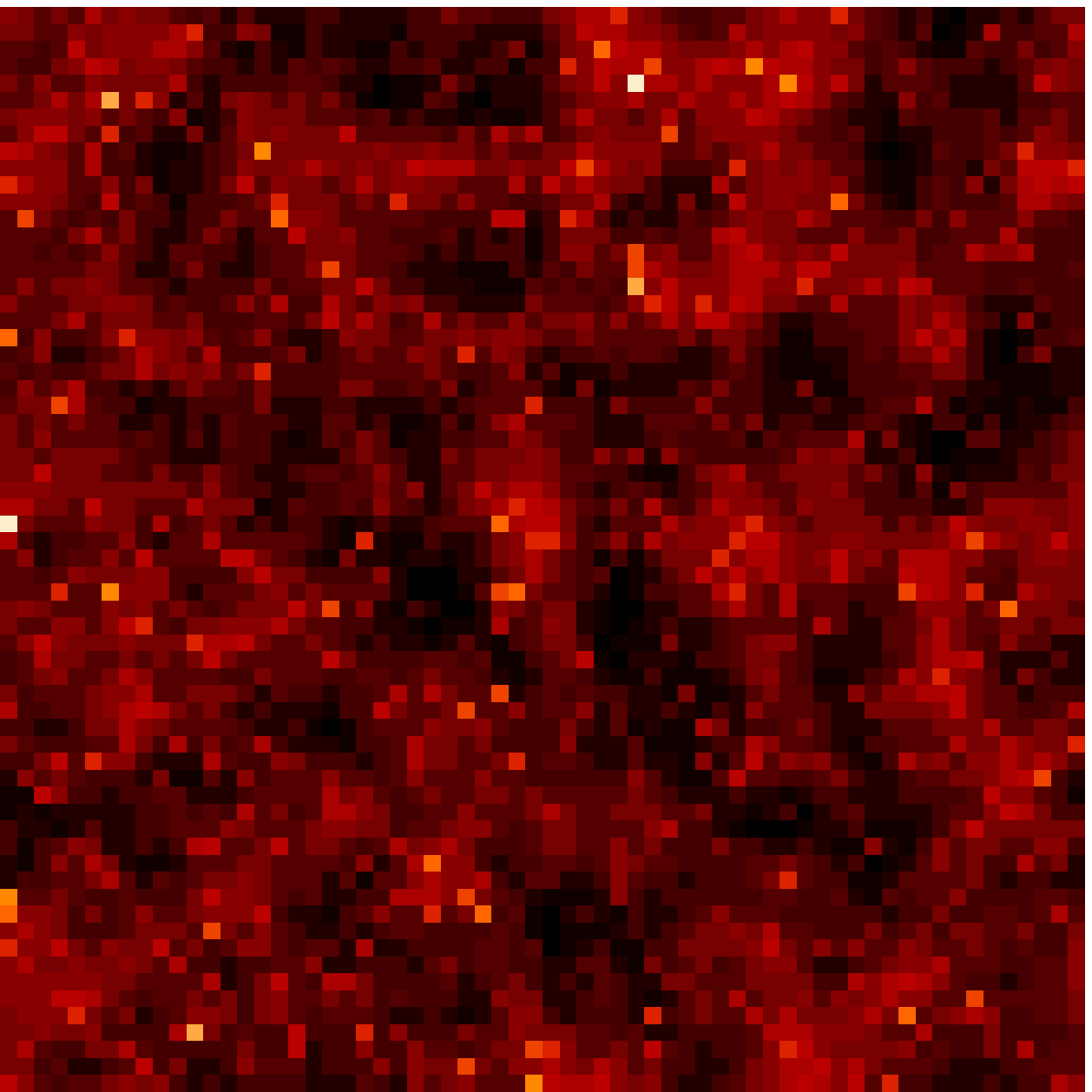,width=50mm}
\hspace{-10mm}
\psfig{figure=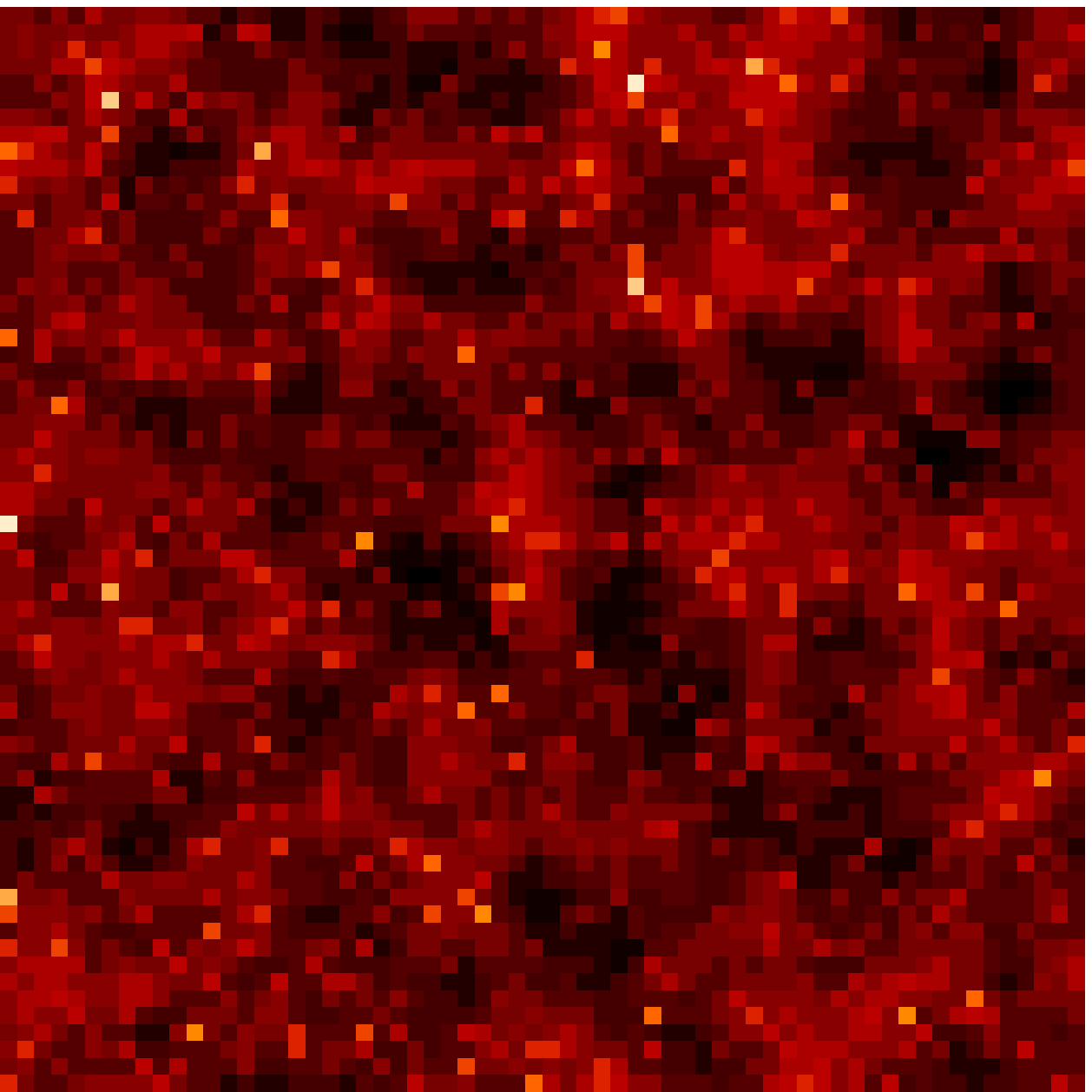,width=50mm}
}
{\hskip 11mm 1 \hskip 39mm 2 \hskip 39mm 3 \hskip 39mm 4}
\vskip 5mm

\centerline{
\psfig{figure=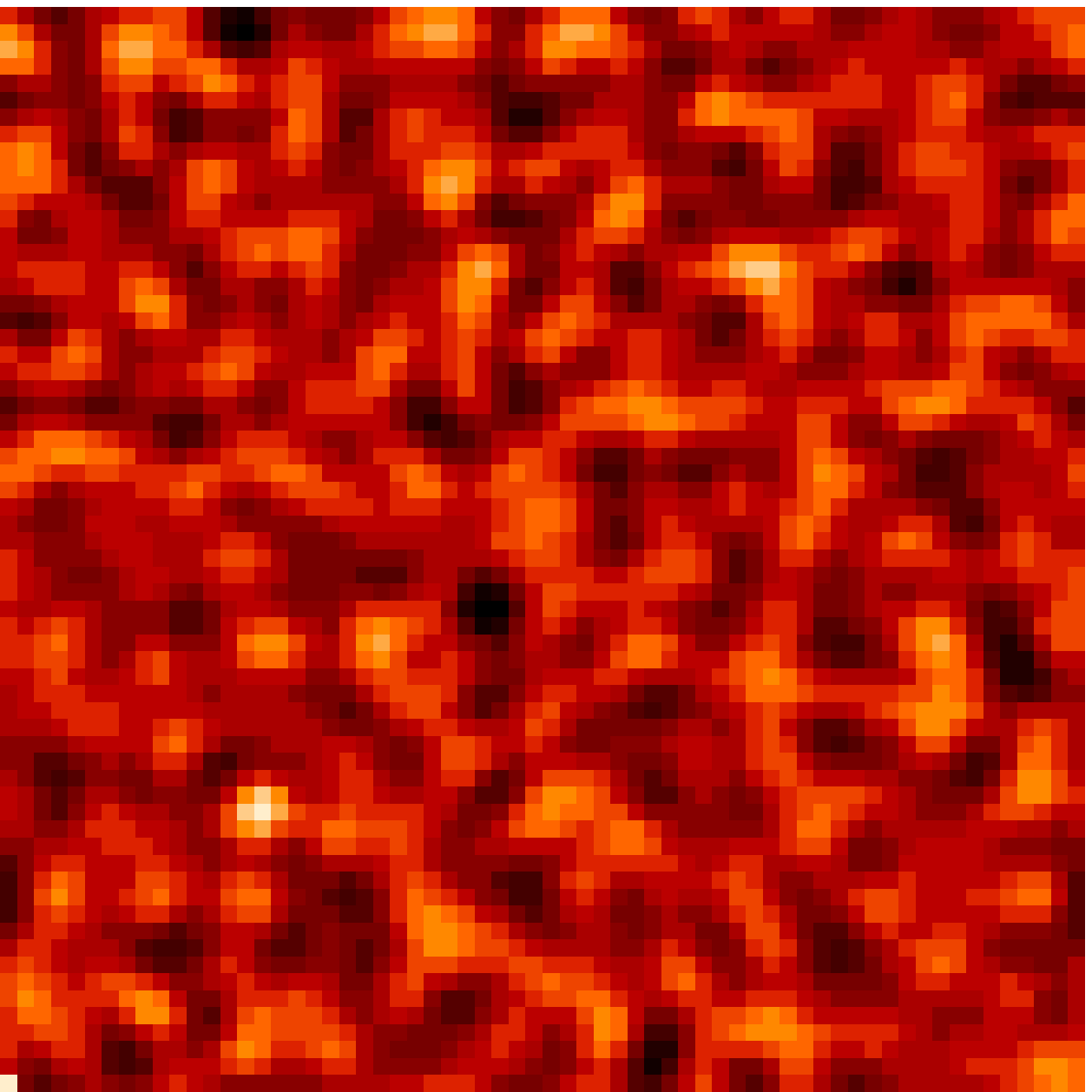,width=50mm}
\hspace{11.5mm}
\psfig{figure=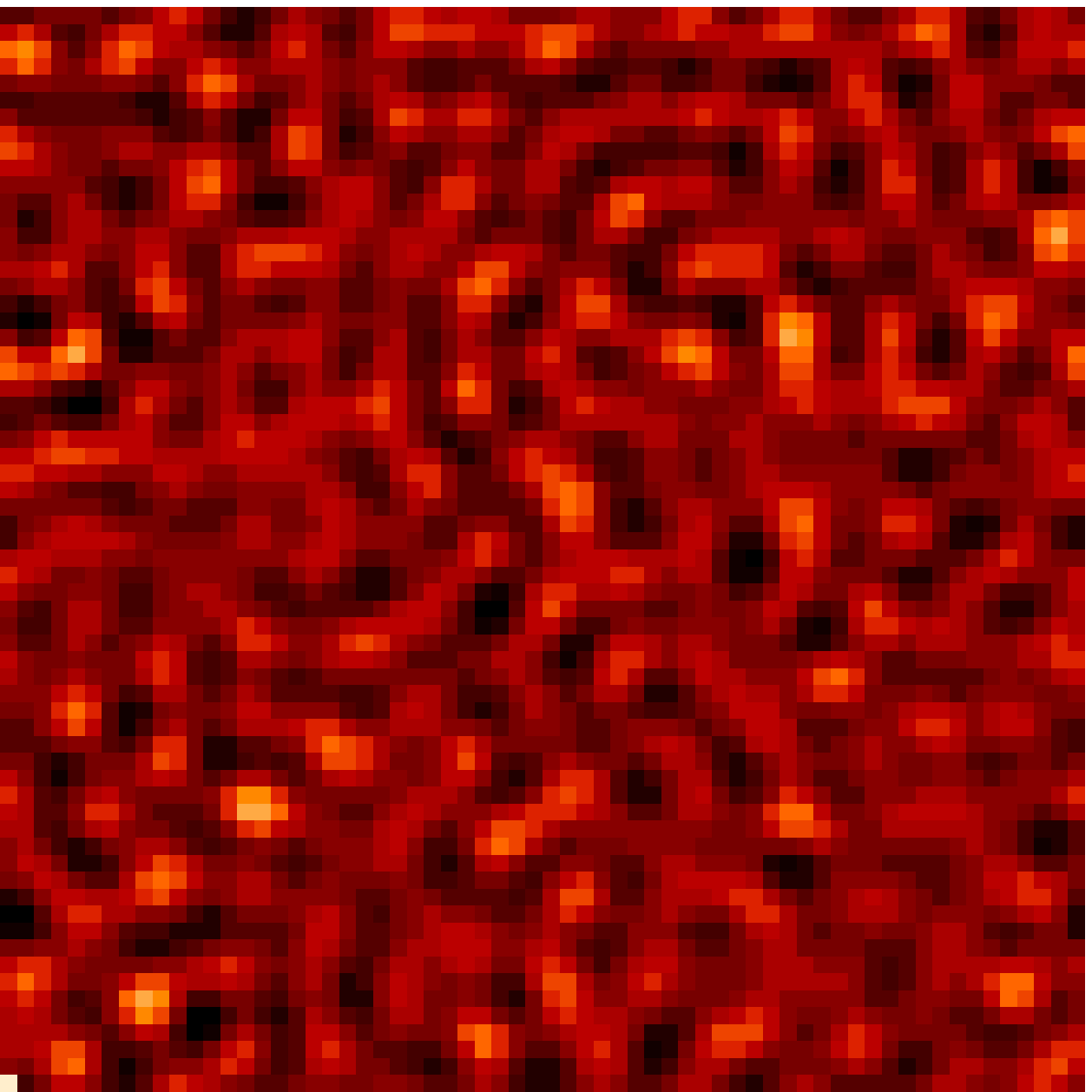,width=50mm}
\hspace{11.5mm}
\psfig{figure=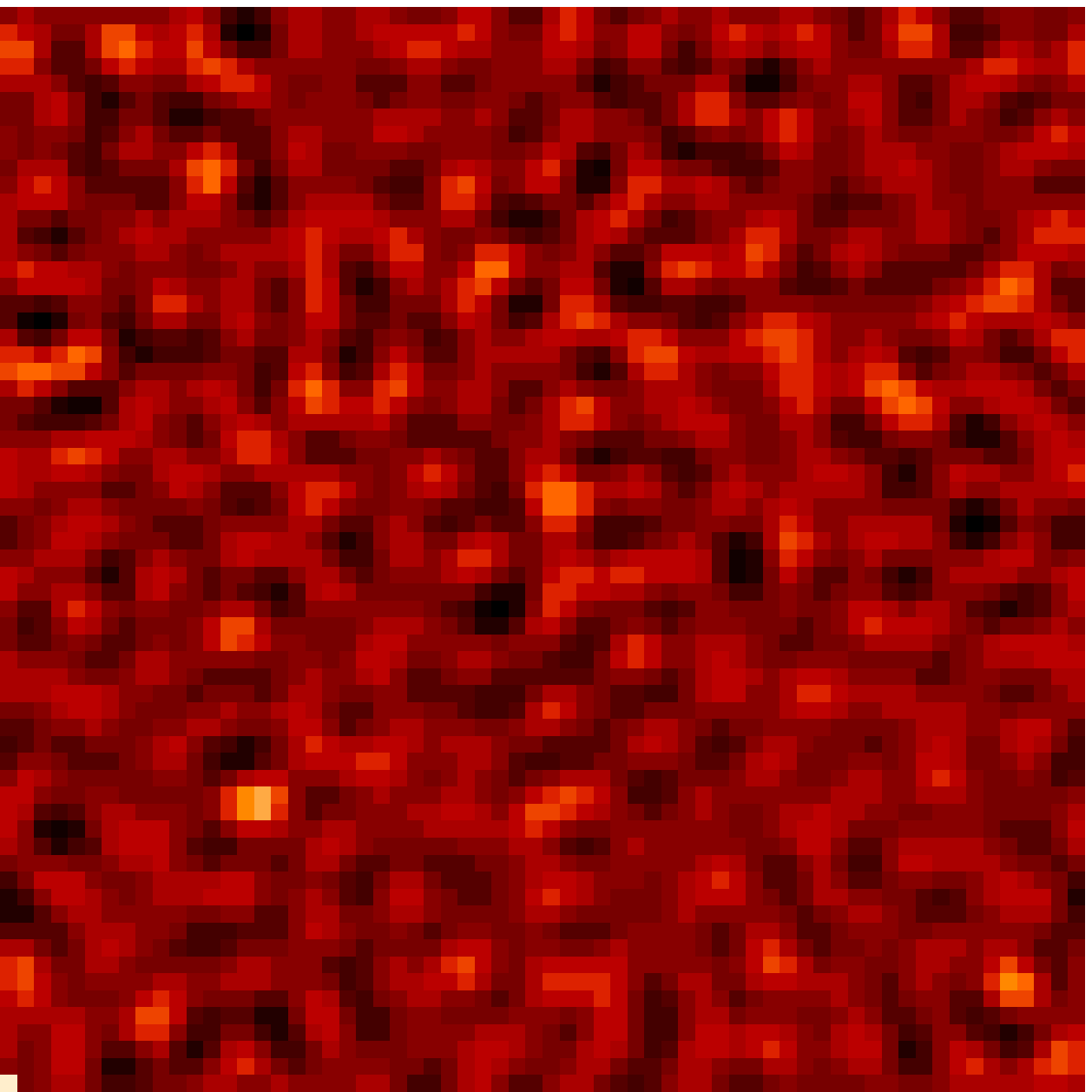,width=50mm}
}
{\hskip 9mm 1-2 \hskip 57mm 1-3 \hskip 58mm 1-4}
%\caption{}
\vskip 5mm
\centerline{\bf Fig.~3}
\end{figure}

\begin{figure}

\centerline{
\psfig{figure=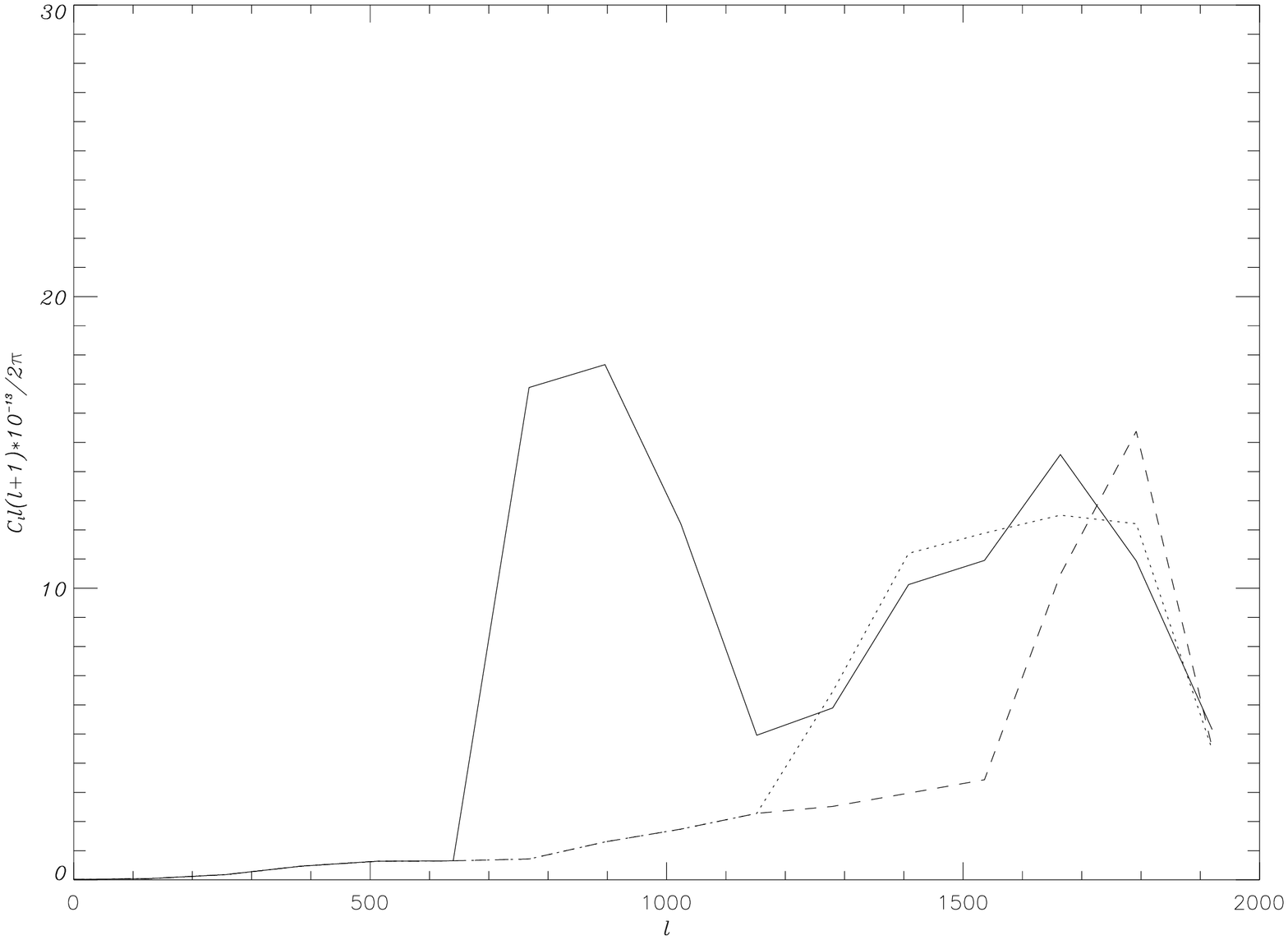,angle=90,width=150mm}
}
%\caption{}
\medskip
\centerline{\bf Fig.~4}
\end{figure}

\section{Acknowledgements}

We would like to thank Yu. Parijskij for interest to
this work.

\noindent This work was partially supported in frame of State Contract
N 40.022.1.1.1106.

\section{References}

\noindent Bajkova, A.T., 2002, astro-ph/0205112.

\noindent Bennett, D.P., Sun Hong Rhie, 1993, ApJ, {\bf 406}, L7--L10.

\noindent Dubrovich, V.K., 1977, Astron. Letters, {\bf 3}, 243 (in
Russian).

\noindent Dubrovich, V.K., 1982, Izvestiya SAO, {\bf 15}, 21 (in Russian).

\noindent Dubrovich, V.K., 1994, A\&A Tr., {\bf 5}, 57.

\noindent Dubrovich, V.K., Lipovka, A.A., 1995, A\&A, {\bf 296}, 301.

\noindent Dubrovich, V.K., 1997, A\&A, {\bf 324}, 27.

\noindent Dubrovich, V.K., 1997a, Pross. HSRA, 22-26 Jan. 1995,
Jodrel Bank, CUP, 189.

\noindent Dubrovich, V.K., 1999, Gravitation and Cosmology, {\bf 5}, 171.

\noindent Dwek, E. et al., 1998, astro-ph/9806129.

\noindent Haarsma, D.B., and Partridge, R.B., 1998, ApJ, {\bf 503}, L5.

\noindent Hanany, S. et al., 2000, Astrophys. J., {\bf 545}, L5.

\noindent Hauser, M.G. et al., 1998, astro-ph/9806167.

\noindent Lange, A.E. et al, 2001, Phys. Rev., D63, 042001.

\noindent Lepp, S., Shull, J. M., 1984, ApJ, {\bf 280}, 465.

\noindent Maoli, R. et al., 1996, ApJ., {\bf 457}, 1.

\noindent Palla, F. et al., 1995, ApJ, {\bf 451}, 44.

\noindent Puy, D. et al., 1993, A\&A, {\bf 267}, 337.

\noindent Sahni, V., Starobinsky, A., 1999, astro-ph/9904398.

\noindent Silk, J., 1994, astro-ph/9405072.

\noindent Stancil, P.D., Lepp, S., and Dalgarno, A., 1996, ApJ, {\bf 458}, 401S.

\noindent Vishniac, E.T., 1987, ApJ, {\bf 322}, 597.

\noindent Zel'dovich, Ya.B., 1978, Sov.Astron.Let., {\bf 4}, 165.

\end{document}